\documentstyle[twocolumn,aps]{revtex}
\input epsf
\newcommand{\authone}[2]{#1 #2,}
\newcommand{\authtwo}[4]{#1 #2 and~#3  #4,}
\newcommand{\auththr}[6]{#1 #2, #3  #4, and~#5 #6,}
\newcommand{\authfour}[8]{#1 #2, #3 #4, #5 #6, and~#7 #8,} 
 
\newcommand{\authmanythr}[6]{#1 #2, #3 #4, #5 #6,} 
 
\newcommand{\private}[2]{ (private communication).}

\newcommand{\yjfm}[5]{  { J. Fluid Mech. }{\bf #2}, #3 (#1).}

\newcommand{\yprl}[5]{  { Phys. Rev. Lett. }{\bf #2}, #3 (#1).}
\newcommand{\ypra}[5]{  { Phys. Rev. A. }{\bf #2}, #3 (#1).}
\newcommand{\ypre}[5]{  { Phys. Rev. E. }{\bf #2}, #3 (#1).}

\newcommand{\yjas}[5]{  { J. Atmo. Sci. }{\bf #2}, #3 (#1).}

\newcommand{\yphy}[5]{  { Physica } {\bf #2}, #3 (#1).}
\newcommand{\yjour}[6]{ { #2} {\bf #3}, #4 (#1).}

\newcommand{\sepl}[2]{  { Europhys. Lett. } (submitted).}
\newcommand{\sana}[2]{  { Astron. Astrophys. } (submitted).}
\newcommand{\tana}[2]{  { Astron. Astrophys. } (to be submitted).}
\newcommand{\pana}[2]{  { Astron. Astrophys. } (to be
published).}
\newcommand{\pmn}[2]{  { Monthly Notices Roy. Astron. Soc. }
(in press).}
\newcommand{\sjfm}[2]{  { J. Fluid Mech. } (submitted).}
\newcommand{\pjfm}[2]{  { J. Fluid Mech. } (to be published).}

\newcommand{\sprl}[2]{  { Phys. Rev. Lett. } (submitted).}
\newcommand{\pprl}[2]{  { Phys. Rev. Lett. } (to be
published).}
\newcommand{\spre}[2]{  { Phys. Rev. E } (submitted).}
\newcommand{\ppre}[2]{  { Phys. Rev. E } (to be published).}
\newcommand{\sprsl}[2]{  { Proc. Roy. Soc. Lond. } (submitted).}
\newcommand{\pprsl}[2]{  { Proc. Roy. Soc. Lond. } (in press).}
\newcommand{\spf}[2]{  { Phys. Fluids } (submitted).}
\newcommand{\ppf}[2]{  { Phys. Fluids } (to be published).}
\newcommand{\spp}[2]{  { Phys. Plasmas } (submitted).}
\newcommand{\ppp}[2]{  { Phys. Plasmas } (in press).}
\newcommand{\sgafd}[2]{  { Geophys. Astrophys. Fluid Dyn. } (submitted).}
\newcommand{\smn}[2]{  { Monthly Notices Roy. Astron. Soc. }
(submitted).}
\newcommand{\sapj}[2]{  { Astrophys. J. } (submitted).}
\newcommand{\papj}[2]{  { Astrophys. J. } (to be published).}
\newcommand{\ssph}[2]{  { Solar Phys. } (submitted).}
\newcommand{\psph}[2]{  { Solar Phys. } (to be published).}

\begin{document}
\title{The energy budget in Rayleigh-B\'enard convection}
\author{R.M. Kerr}
\address{NCAR, Boulder, CO 80307-3000;
Atmospheric Sciences, University of Arizona, Tucson, AZ  85721-0081}
\maketitle
\begin{abstract}
It is shown using three series of Rayleigh number simulations 
of varying aspect ratio $AR$ and Prandtl number $Pr$ that the normalized
dissipation at the wall, while significantly greater than 1, 
approaches a constant dependent upon $AR$ and $Pr$.  It is also found
that the peak velocity, not the mean square velocity, obeys
the experimental scaling of $Ra^{0.5}$.  The scaling of the mean square velocity
is closer to $Ra^{0.46}$, which is shown to be consistent with
experimental measurements and the numerical results for the
scaling of $Nu$ and the temperature if there are strong correlations
between the velocity and temperature.  
\end{abstract}
% \section{Introduction}
This Letter will analyze the energy budget
in three-dimensional simulations of Rayleigh-B\'enard convection
with the objective of testing theoretical assumptions used to explain
the laboratory observation 
of non-classical heat flux (Nusselt number $Nu$) exponents in 
classical turbulent Rayleigh-B\'enard \cite{Heslotetal87,Niemelaetal00}.
The budget equation for the average kinetic energy $q^2/2$ as a function
of height $z$ \cite{DeardorffWillis67} is
\begin{equation}
{1\over2}{\partial\over{\partial t}} \overline{q^2} = - \overline{pw}_{,z}
-\overline{0.5wq^2}_{,z}
+ Ra\overline{w \theta} + Pr\partial^2_z \overline{q^2} -Pr\epsilon(z)
\label{eq:ebudget}
\end{equation}
where shear production $-\overline{uw(dU/dz)}$ is part of the
turbulent production term $-\overline{0.5wq^2}_{,z}$.  
Based upon the observation
of strong shears in the boundary layer \cite{Werne,ZocciMLib} 
and large-scale flows \cite{RubyHoward81}, it
was first assumed that
the dissipation is primarily concentrated in the boundary layer
and is turbulent \cite{SS90}. More recently it has been suggested that
the boundary layer is laminar and the
distribution of the total energy dissipation 
is Rayleigh number dependent\cite{G-Lohse00}.  Both of these
approaches predict crossovers in scaling behavior.
Another theory \cite{Castaingetal89} makes mixing layer assumptions
and does not predict these crossovers.  The objective in this
Letter is to look at the basis for these assumptions using
numerical simulations.

Assuming that $Nu=\overline{w\theta}/\kappa d\Theta/dz\sim Ra^{\beta_T}$, 
the classical exponent is $\beta_T=1/3$, with suggestions 
since the mid-60's \cite{Roberts65,Herring66}
that there might be corrections to this exponent.
The original experimental result \cite{Heslotetal87} showing that there
are significant corrections has recently
been extended and refined to give $Nu\sim Ra^{.309}$ over nearly 
10 decades of Rayleigh number $Ra$ \cite{Niemelaetal00}.  
However, detailed experimental information
such as budgets cannot be obtained from these large $Ra$ experiments.
The only information available besides temperature statistics at a single
point is that the Reynolds number based upon vertical velocity fluctuations
$w$ midway up a sidewall 
goes as $Re=wd/\nu\sim Ra^{1/2}$.  These two observations are incompatible 
with standard turbulent parameterizations
because the total dissipation is constrained 
to be $\epsilon_T=Ra(Nu-1)\sim Ra^{1.31}$, while
the standard turbulent prediction of $\epsilon_T=Re^3$ 
gives $\epsilon_T\sim Ra^{1.5}$.

One reason a laminar boundary layer has been suggested
is that this incompatibility can be resolved if the laminar relationship 
for dissipation is used, 
$\epsilon_T\sim Re^{5/2}\sim Ra^{1.28}$.
This would suggest that convective flows are not filled 
with cascading eddies, but are instead filled with 
strong local laminar shears, including a laminar
boundary layer.  One way to possibly determine whether the
boundary layer is laminar or turbulent is to
divide the total energy dissipation $\epsilon_T$ at
some arbitrary boundary layer thickness $\lambda_{BL}$ into 
the dissipation in the boundary layers $\epsilon_{BL}$ and
the dissipation in the bulk $\epsilon_{B}$.  A new theory for
convective scaling \cite{G-Lohse00} has assumed that in a turbulent
boundary layer that $\epsilon_{BL}$ will be the same order or
greater than $\epsilon_B$,  while if the boundary layer is
laminar then $\epsilon_{BL}/\epsilon_B$ will decrease
as $Ra$ increases.  This is justified using the scaling
laws above for dissipation in laminar and turbulent boundary layers.

Since the average (or total) dimensionless dissipation
$\epsilon_T=Ra(Nu-1)$ is increasing with $Ra$, it is possible
that even if the boundary layer is laminar at one $Ra$, that an
$Ra$ could be reached where the boundary layer becomes
unstable and turbulent. Then for higher $Ra$, 
as in a shear driven turbulent boundary layer,
$\epsilon_{BL}/\epsilon_B\longrightarrow$constant would appear.

One way to obtain detailed information and test these properties 
is to use simulations.
Simulations have reproduced most
of the scaling laws and statistical properties of the higher
Rayleigh number experiments \cite{Kerr96,KerrH00}, including
$\beta_T$ between 2/7 and 1/3.  One of the properties that
simulations can determine are thermal and velocity boundary layer
thicknesses.  The thermal boundary layer
thickness $\lambda_T$ can be determined using either $1/Nu$ or the
position of the peak of temperature fluctuations $\overline{\theta^2}$. 
Based upon experience from analysis of classical shear driven boundary layers,
three definitions of the velocity boundary layer thickness 
$\lambda_{BL}$ that can be calculated for convection use the velocity to give
$z^*=1/Re$, use the wall shear stress $\tau=\overline{\partial u/\partial x}$
to give $z^+=\sqrt{\tau}$, and use the position
of the peak of the horizontal velocity fluctuations $\lambda_u$. 
The dimensionless forms have been used in these definitions.  

These and additional definitions of the boundary layer thickness could
be used to define the dissipation in the boundary layer $\epsilon_{BL}$,
which we would like to relate to the dissipation at the wall.
While $\epsilon_{BL}$  depends
upon the definition of the thickness of the boundary layer, a
relationship that could hold for all the definitions above is
\begin{equation}
\epsilon_{BL}/\epsilon_T \leq \big(\epsilon_W\lambda_{BL}/(d\epsilon_T)\big)
\label{eq:ebnd}
\end{equation}
where $\epsilon_W$ is the dissipation at the wall.  This relationship
is not very restrictive, but if $\epsilon_W/\epsilon_T$ were bounded,
because $\lambda_{BL}$ is a decreasing function 
of $Ra$ for all the definitions given above,
$\epsilon_{BL}/\epsilon_T$ would be bounded.  

In a classical
shear-driven boundary layer, all of the definitions
of $\lambda_{BL}$ given above scale with $Re$ in the same manner.
A major difference in convection simulations\cite{Kerr96,KerrH00}
is that each definition of $\lambda_{BL}$ 
has a separate power law dependence on $Ra$.  That is, if 
\begin{equation} 
\lambda_u\sim Ra^{-\beta_u}\quad{\rm and}\quad 
z^+=\sqrt{\tau}\sim Ra^{-\beta^+}
\label{eq:betauz}\end{equation}
then $\beta_u\approx1/7$ and $\beta^+>\beta_T$ has been found \cite{Kerr96}.
Because these results and the theories all indicate that $\lambda_{BL}$
is decreasing rapidly with $Ra$, by applying (\ref{eq:ebnd}) 
both the turbulent and the laminar boundary layer theories predict
that $\epsilon_W/\epsilon_T$ should increase as $Ra$ to a power law.

Three simulations have been analysed 
to determine the dependence of $\epsilon_W/\epsilon_T$ upon $Ra$.
The cases to be discussed are $AR=4$ and $Pr=0.3$ for $10^5<Ra<10^7$, 
$AR=4$ and $Pr=7$ for $10^4<Ra<10^7$ from earlier work \cite{KerrH00}, 
and $AR=1$ and $Pr=0.7$ for $10^6<Ra<8\times10^7$ where
Prandtl number $Pr=\nu/\kappa$ and aspect ratio $AR=width/height$.
The numerics \cite{KerrH00} are pseudospectral, using Chebyshev
polynomials in the vertical to provide more resolution and
no-slip, constant temperature boundary conditions at top and bottom walls. 
In the horizontal, sines and cosines are used
to represent free-slip, insulating walls.  Profiles will be
shown only for the $AR=1$ case to save space and because this case is new.
% $Pr=0.3$ was chosen as the low end because it
% seems to show a transition from low $Pr$ behavior with $Nu~\sim~Ra^{1/4}$ 
% to stronger growth with $Ra$ within the range of simulated $Ra$ and
% because for the more detailed statistics presented here the highest
% $Ra$ calculation from before \cite{KerrH00} for $Pr=0.07$ was not
% adequately resolved. 
% The lowest $Ra$ used for the averages for $Pr=7$
% is $Ra=5\times10^5$ because $Ra=10^4$ is not turbulent.

\epsfxsize=8.5cm
\begin{figure}[htbp] 
\epsfbox{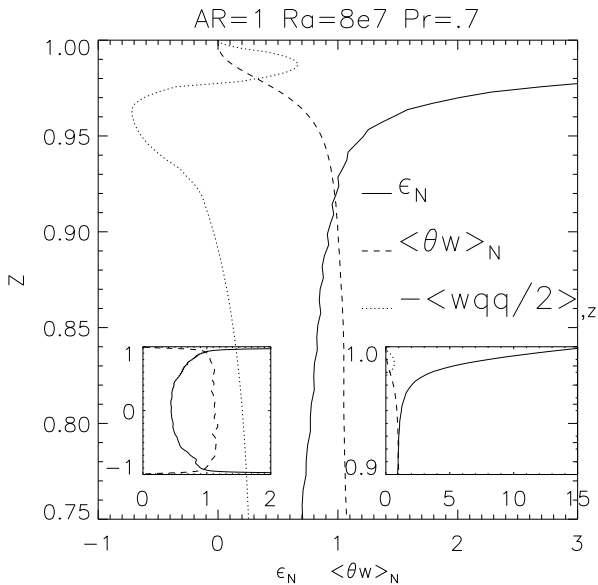}
\caption[]{Production, dissipation and turbulent transport of kinetic
energy from (\ref{eq:ebudget})
for the simulation $AR=1$, $Pr=0.7$ and $Ra=8\times10^7$.  
All dimensionless terms are normalized by $RaNu$ so that the production across
the center $\overline{w\theta}_N$, which is also the heat flux, is 1.  
Insets show production and dissipation through the entire box 
(height $d=2$) and very near the wall, where dissipation is very large.
The difference between production and dissipation 
$\overline{w\theta}(z)-\epsilon(z)$ is the total transport term.
As discussed, the
pressure transport is close to the difference between the production
and dissipation and so is much larger than the turbulent transport
$\overline{wq^2/2}_{,z}$.}
\label{fig:transz}
\end{figure}
Previous work \cite{Kerr96} demonstrated that $\epsilon_W$
was much larger than the average dissipation
and that $\epsilon_W/\epsilon_T$ was increasing with $Ra$.
Fig. \ref{fig:transz} shows some of
the terms in eq. (\ref{eq:ebudget}) for $AR=1$, $Pr=0.7$
and $Ra=8\times10^7$, all normalized by $RaNu$. For example,
$\epsilon_N(z)=\epsilon(z)/RaNu$.
The $AR=4$ cases are similar except that $\epsilon_{WN}=\epsilon_W/RaNu$ 
is smaller,
$\min(\epsilon_N(z))$ is closer to 1, and $\overline{wq^2/2}_{,z}$ 
is even smaller than here.

\epsfxsize=8.5cm
\begin{figure}[htbp] 
\epsfbox{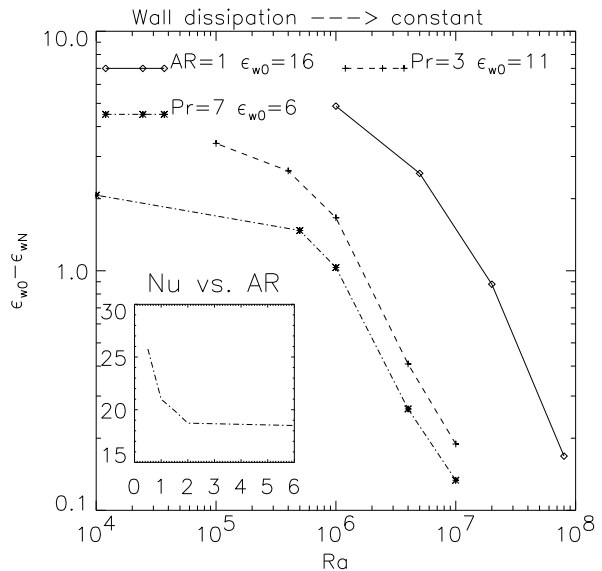}
\caption[]{$\epsilon_{W0}(AR,Pr)-\epsilon_N$ vs. Ra.  
The decrease is approximately
$Ra^{-0.8\pm0.1}$ for all three cases (neglecting
the lowest $Ra$ for $Pr=0.3$ and 7).  The values of $\epsilon_{W0}(AR,Pr)$ 
used are given.
The inset shows the dependence on aspect ratio of $Nu$ for $Ra=10^7$
and $Pr=0.7$.}
\label{fig:disswinv}
\end{figure}
While this analysis has confirmed that $\epsilon_W$
is much larger than the average dissipation
and $\epsilon_W/\epsilon_T$ is increasing with $Ra$, careful
examination revealed that there appears to be an upper
bound to $\epsilon_W/\epsilon_T$, which will be denoted $\epsilon_{W0}$
and is found to be strongly dependent upon $Pr$ and $AR$.
Fig. \ref{fig:disswinv} 
shows the normalized dissipation at the wall plotted as
$\epsilon_{W0}-\epsilon_{WN}$.
For all three cases, $\epsilon_{W0}-\epsilon_{WN}~\sim~Ra^{-0.8\pm0.1}$.  
While the choice of this particular form is subjective, the consistency in
$\epsilon_{WN}\longrightarrow \epsilon_{W0}(AR,Pr)$ as $Ra$ grows for all
$AR$ and $Pr$ appears to be robust and contradicts the predictions of both the
turbulent and laminar boundary layer theories.  
Due to the bound (\ref{eq:ebnd}) on $\epsilon_{BL}$,
then $\epsilon_{BL}/\epsilon_B\rightarrow0$
as $Ra\rightarrow\infty$ at least as fast as
\begin{equation}
\epsilon_{BL}/\epsilon_B<\epsilon_{W0}\lambda_{BL}\sim Ra^{-1/7}
\label{eq:BL0}
\end{equation}
where $-1/7$ comes from using $\lambda_{BL}=\lambda_u$ from (\ref{eq:betauz}).

If this trend were to continue to higher $Ra$ it would imply that
the boundary layer should not
be characterized as either a laminar or a turbulent shear-driven 
boundary layer as has been assumed up to now.  Furthermore, it
implies that the smallest
length scales in the problem are multiples of the Kolmogorov scale
$\eta=\epsilon_T^{-1/4}=(Ra(Nu-1))^{-1/4}$.  For example,
the wall boundary layer thickness taken from the wall shear stress
$z^+=\sqrt{\tau}$ should scale as $\eta$. If $\beta_T=0.309$, then
$\beta^+=0.327$ is predicted.

Now let us consider the mechanisms responsible for transferring
energy from the bulk to the boundary layer. This is necessary because
Fig. \ref{fig:transz} shows that the production of kinetic
energy, which is equivalent to the convective heat flux
$\overline{w\theta}$, is found only in the center of the box, but
the peak of the dissipation is at the wall.  
For there to be a turbulent boundary layer, there would have to 
be large turbulent production or shear production terms.
With the normalization used in Fig.  \ref{fig:transz}, 
$\overline{w\theta}/(RaNu)\approx 1$ in the center and 
$\epsilon(z)/(RaNu)$ is slightly less than 1, 
which is compensated for by extra dissipation in the boundary layer.

The three transport terms in (\ref{eq:ebudget}) that transfer
energy from the bulk to the boundary layer are
diffusive transport $Pr\partial^2_z q^2 $, the pressure transport 
$-\overline{pw}_{,z}$ and the turbulent transport
$-\overline{0.5wq^2}_{,z}$.  $Pr\partial^2_z q^2 $ and
$-\overline{0.5wq^2}_{,z}$ can be calculated directly while
$-\overline{pw}_{,z}$ can be calculated from the difference between
all the remaining terms.  
Diffusive transport (not shown) is found to be large
only very near the wall, for $1-z<0.02$ in Fig. \ref{fig:transz}.
For $1-z<0.1$ the turbulent transport is much less than the
difference between the production and the dissipation, and therefore
the turbulent transport is much less than the pressure transport term.
This would be consistent with the dissipation in the boundary layer
being much less than either the predictions of a turbulent or a laminar
boundary layer, both of which depend upon shear production, which
is part of the turbulent production term.

The observation that the pressure transport dominates is not new.
In the geophysical literature over a wide
range in $Ra$ beginning at low $Ra$ in the laboratory
\cite{DeardorffWillis67} and extended to higher $Ra$ in atmospheric observations
\cite{LemoneGATE,Lenschowetal80}, it is found that
the primary source of kinetic energy in the boundary layer is
the pressure transport term.  

If the boundary layer is neither a shear driven turbulent nor
a laminar boundary layer, then what mechanism is responsible for the
observed scaling of the velocity scale or Reynolds number $Re=wd/\nu$?  
Perhaps the origin is in the details of plume dynamics.  It has
previously been found \cite{Yakhot89}, and confirmed by these
simulations, that there is a nearly perfect correlation in the
bulk between the vertical velocity $w$
and temperature fluctuations $\theta$, where the total temperature
is $T(x,y,z)=\overline{T}(z)+\theta(x,y,z)$.  What is found everywhere in these
calculations, except near the wall ($z<\lambda_T=d/Nu$), is that 
\begin{equation}
\overline{\theta w}\approx0.9\overline{w^2}^{1/2}\overline{\theta^2}^{1/2}
\label{eq:wT}
\end{equation}
This would be consistent with visualizations that show small plumes
dominating in the boundary layer \cite{ZocciMLib,Kerr96,KerrH00}.

\epsfxsize=8.5cm
\begin{figure}[htbp] 
\epsfbox{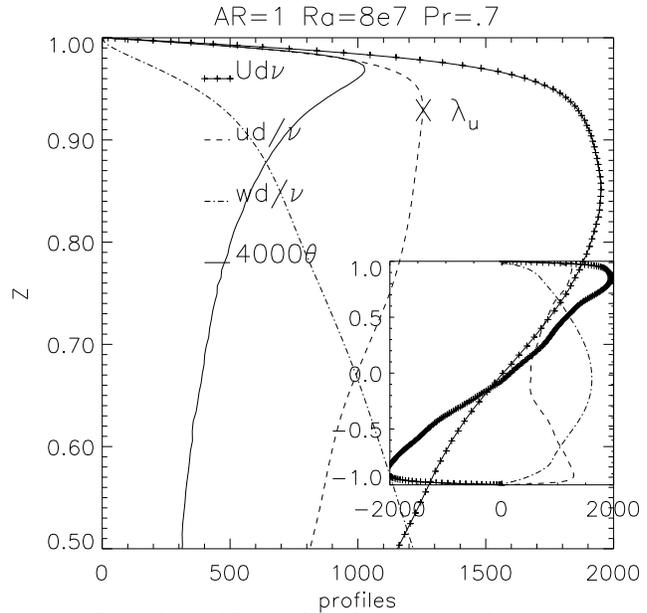}
\caption[]{Reynolds numbers based upon the mean horizontal velocity
in the direction of maximum shear $Vd/\nu$, upon the horizontal
fluctuations in velocity $ud/\nu=(\overline{u^2+v^2})^{1/2}d/\nu$, upon 
the vertical velocity fluctuations $wd/\nu$, and the temperature
fluctuation variance $\overline{\theta^2}^{1/2}$. The inset shows the
profiles over the entire domain.}
\label{fig:ar1prof}
\end{figure}
Using (\ref{eq:wT}), it can be predicted that in the center that if 
$Nu=\overline{\theta w}/\kappa(d\Theta/dz)\sim Ra^{\beta_T}$, 
$\theta'={\overline{\theta^2}^{1/2}}\sim Ra^{-\delta_c}$,
and $Re={\overline{w^2}^{1/2}}/\nu d\sim Ra^{\gamma}$ that
\begin{equation} \gamma=\beta_T+\delta_c.  \label{eq:deltac}\end{equation}
The scaling for the simulations and 
experiments discussed here is given in Table \ref{table1}.  
For the experimental values
\cite{Niemelaetal00} of $\beta_T=0.309$ and
$\delta_c=0.145$, then (\ref{eq:deltac}) would
predict $\gamma=0.454$, which is the origin of 
the value given in Table 1.  However,
what is found (private communication)
for the most recent experiment \cite{Niemelaetal00}
and most earlier experiments \cite{Chavanneetal97}
% \cite{Castaingetal89,Xinetal96,Chavanneetal97,QiuXia98}
is $\gamma=0.5$.  
For $AR=4$, $Pr=0.3$ and $AR=1$, $Pr=0.7$: $\gamma=0.46$
and for $AR=4$, $Pr=7$: $\gamma>0.5$.  

These inconsistencies can be resolved
by looking at the scaling of the maximum vertical velocity in the
simulations, where it is found that $Re_{max}\sim Ra^{0.5}$ for
all three simulated cases.  Furthermore, this maximum is always found roughly 
midway up a sidewall in a strong persistent plume.  Midway
up a sidewall is also where the experimental measurements
of velocity are taken.  Therefore, this indicates that these
experimental velocity measurements are not representative of the velocity
as a whole, but only representative of the velocity at the walls that probably
comes from the maximum possible vertical velocity within plumes.  
The reason $AR=4$, $Pr=7$ gives $\gamma>0.5$
can be understood by noting that even at its
highest $Ra$ of $10^7$, the $Re$ for this case is not turbulent
and that in the published visualizations \cite{KerrH00} this case
is dominated by laminar plumes.  Therefore, the velocity in individual
plumes seems to obey $\gamma_{max}=0.5$, but when the flow is 
turbulent the average exponent is closer to $\gamma=0.46$.
This could represent an average between $\gamma_{max}=0.5$ and
the exponent found for the Reynolds number dependence of the
large-scale circulation, where $\gamma=0.43$ \cite{AshkenaziSteinberg99}.
% For the other cases, at their highest $Ra$, properties such
% as the normalized vortex stretching \cite{Kerr96}
% are consistent with at least the interior being turbulent.
\begin{table}
% \begin{center}
\begin{tabular}{lcccc}
% \hline
Expon  & Expt. & $AR=1,Pr=0.7$ & 4,0.3 & 4,7 \\
$\beta_T$ & 0.309 & 0.27$\pm$0.02&0.31$\pm$0.02&0.30$\pm$0.02\\
$\gamma_{\max}$ & 0.50 & 0.50$\pm$0.01&0.50$\pm$0.01&0.51$\pm$0.01\\
$\delta_c$ & 0.145 & 0.17$\pm$0.02&0.14$\pm$0.02&0.16$\pm$0.03\\
$\gamma$ & 0.454 & 0.46$\pm$0.02&0.46$\pm$0.01&0.52$\pm$0.03\\
$Re(Ra=10^7)$ & & 434 & 3000 & 14 \\
\end{tabular}
\caption{Experimental [2] %\cite{Niemelaetal00} 
and numerical exponents.  Aspect ratio and Prandtl number for the
three numerical cases is given.  The experimental value for velocity
scaling is given as $\gamma_{\max}$. How the experimental value
for $\gamma$ is gotten is explained in the text.  Errors for the
simulations are based upon scatter of exponents taken between $Ra$.
$Re=ud/\nu$ is taken at $Ra=10^7$ for all three cases.} \label{table1}
% \end{center}
\end{table}

It has been found that the dissipation at the wall approaches a multiple
of the mean dissipation across the box, contrary to the assumptions
of the effect of shear upon the thermal boundary layer.  Therefore,
it does not seem that the dynamics in the boundary layer can be
governed by shears, whether they be laminar or turbulent.  Instead,
it is suggested that plumes might dominate the dynamics and perhaps
a new model of turbulent convection
should be constructed based upon plume dynamics.  The basis
for this suggestion is how the velocity scales.
Evidence is presented that the experimentally
observed scaling of the Reynolds number as $Re\sim Ra^{\gamma}$,
with $\gamma=0.5$ can be reproduced by the simulations
only if this is taken to be the maximum vertical
velocity in persistent plumes along the sidewalls. So one
should replace $\gamma$ by $\gamma_{\max}$ in the experiments.
Based upon the simulations, it is proposed that 
the simulated exponent based upon the average kinetic energy
is $\gamma\approx 0.46$, which is shown to be consistent with
the experimental measurements of the scaling of the temperature
fluctuations and $Nu$.  This suggests that
it would be useful to have several determinations 
of the Reynolds numbers from a single experiment.  That is,
Reynolds numbers based upon 
velocity measurements taken near a sidewall, in the interior, 
and from the large-scale circulation.  

NCAR is supported by the National Science Foundation.  Conversations
with K.R. Sreenivasan, D. Lohse, C. Doering, B. Shraiman, J.R. Herring 
and others while the author was a participant in the Program on
Hydrodynamic Turbulence at the Institute for Theoretical Physics,
University of California, Santa Barbara are acknowledged.

\end{document}